\documentclass{optica-article}

\journal{opticajournal} 

\articletype{Research Article}
\usepackage{lineno}

\usepackage{algorithm}
\usepackage{algpseudocode}
\usepackage{siunitx}

\newcommand{\todo}[1]{}
\renewcommand{\todo}[1]{{\color{red} TODO: {#1}}}
\begin{document}

\title{Wave optical model for tomographic volumetric additive manufacturing}

\author{Felix Wechsler$^{1,\ddagger}$, Carlo Gigli\authormark{1}, Jorge Madrid-Wolff\authormark{1}, and Christophe Moser\authormark{1}}

\address{\authormark{1}Laboratory of Applied Photonics Devices, École polytechnique fédérale de Lausanne, Lausanne, Switzerland}

\email{$^\ddagger$info@felixwechsler.science} 

\begin{abstract*} 
Tomographic Volumetric Additive Manufacturing (TVAM) allows printing of mesoscopic objects within seconds or minutes.
Tomographic patterns are illuminated onto a rotating glass vial which contains a photosensitive resin.
Current pattern optimization is based on a ray optical assumption which ultimately leads to limited resolution around \SI{20}{\micro\meter} and varying throughout the volume of the 3D object.
In this work, we introduce a rigorous wave-based optical amplitude optimization scheme for TVAM which shows that high-resolution 
printing is theoretically possible over the full volume.
The wave optical optimization approach is based on an efficient angular spectrum method of plane waves with custom written memory efficient gradients and allows for optimization
of realistic volumes for TVAM such as $(\SI{100}{\micro\meter})^3$ or $(\SI{10}{\milli\meter})^3$ with $550^3$ voxels and 600 angles. Our simulations show that ray-optics start to produce artifacts when the desired features are \SI{20}{\micro\meter} and below and more importantly, the amplitude modulated TVAM can reach micrometer features when optimizing the patterns using a full wave model.
\end{abstract*}

\section{Introduction}
\begin{figure}[ht]
    \centering
    \includegraphics[width = .9\textwidth]{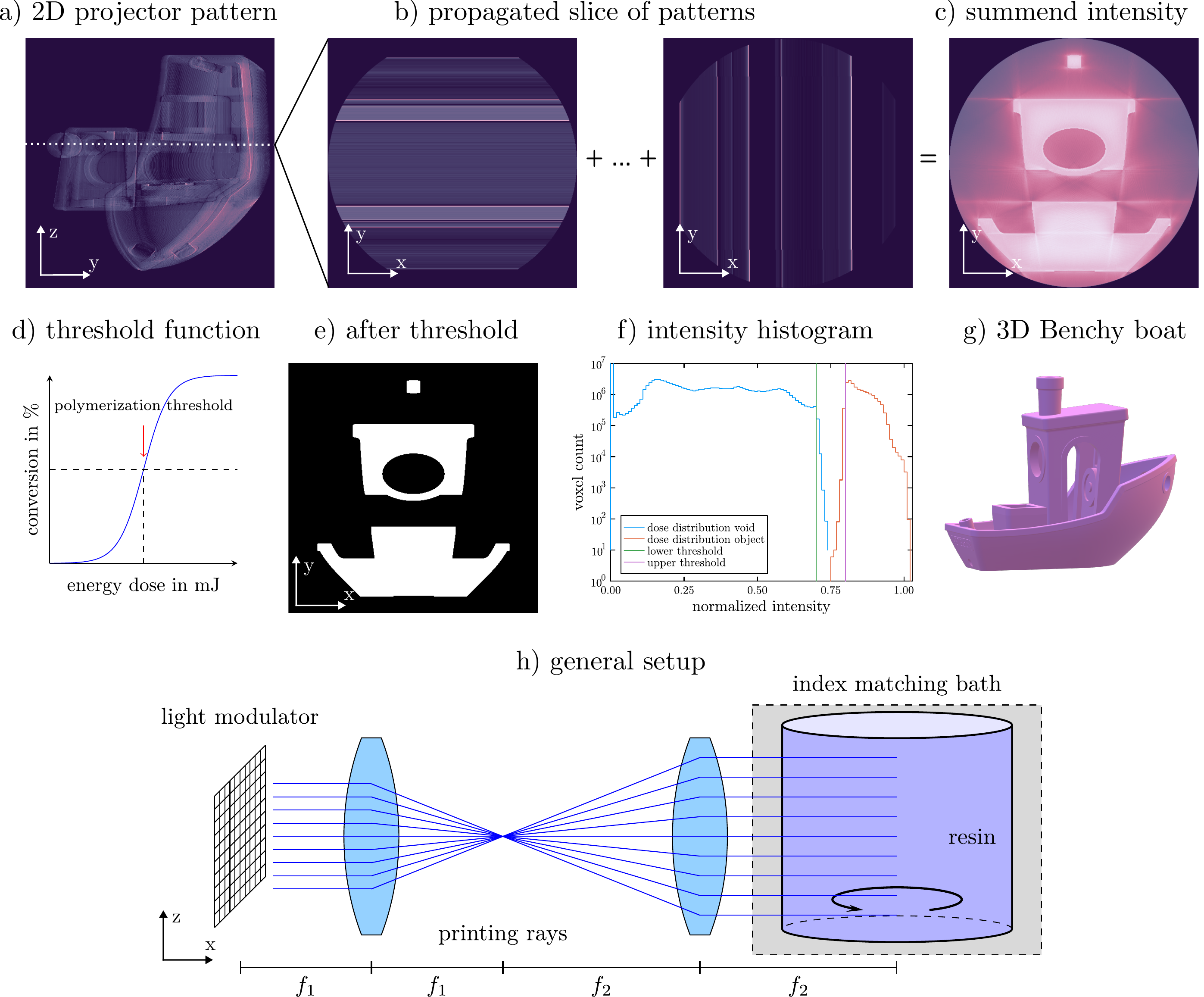}
    \caption{The general principle behind TVAM. a) a set of 2D projection patterns is propagated into space. b) shows how a slice of the pattern propagates through the volume and c) how the incoherent sum results in a total energy dose. d) the object polymerizes if it reaches an energy threshold. 
        e) polymerization threshold results in a printed slice. f) is the intensity histogram of b). g) is the 3D view of the Benchy boat. h) is the general setup.}
    \label{fig:general}
\end{figure}

Tomographic Volumetric Additive Manufacturing (TVAM) \cite{Kelly_Bhattacharya_Heidari_Shusteff_Spadaccini_Taylor_2019, bernal2019volumetric} is a 
novel technique allowing for rapid 3D printing.
As shown in \autoref{fig:general}h), it is based on a computed tomography principle (CT) which consists of projecting 2D light patterns from many angles into a photosensitive resin.
The light patterns build up a cumulative 3D energy dose distribution within the photosensitive resin. 
Above a specific light dose received at a point in space, enough conversion of the monomer into polymer occurs, thereby solidifying the resin. To compute this set of 2D light patterns, most recent works have built upon the Radon transform \cite{kak2001principlesCompTomo}. A common approach is to voxelize the 3D model, calculate the tomographic projections using an implemented Radon transform function, and then filtering these projections. The Radon transform assumes un-attenuated, un-distorted, straight ray propagation, which differs from the physical implementation. The naive approach can be corrected to compensate for lack of telecentricity and limited etendue \cite{webber2023versatile}.
To include absorption of the ray upon propagation, the attenuated inverse Radon transform \cite{FNatterer_2001} is the correct model.
\autoref{fig:general} shows an example of the computed tomographic patterns to construct the Benchy Boat\footnote{\url{https://www.3dbenchy.com/}} for a slice. The object shown in Figure 1e), after thresholding, consists of solidified pixels. 

Finding the optimal set of patterns for the projector has been a problem that is being tackled since the introduction of TVAM. Gradient descent-based methods have been proposed in several different works \cite{Kelly_Bhattacharya_Heidari_Shusteff_Spadaccini_Taylor_2019, Loterie_Delrot_Moser_2020, Bhattacharya_Toombs_Taylor_2021}, with various loss functions used to achieve different properties for the prints. 
The object-space optimization algorithm proposed by Rackson \textit{et al.} \cite{Rackson_Champley_Toombs_Fong_Bansal_Taylor_Shusteff_McLeod_2021} is a conceptually fast algorithm 
and can be implemented without explicit gradient descent formulation. The forward model in this work is the attenuated Radon transform.
Orth \textit{et al.} presented a deconvolution approach for TVAM\cite{Orth_Webber_Zhang_Sampson_de} that accounts for diffraction and for the effect of chemical diffusion of polymerization inhibitors as a convolutional kernel in the optimization scheme. Although this method introduces a diffraction model,  it is still based on ray optics by treating a single ray, as not spatially sharp but instead describing it as a Gaussian intensity function over a spatial region.

Further, a general loss function for TVAM has been proposed to incorporate material specific effects or grayscale dose into the model\cite{li2023tomographic}. 
The same work also mentions the possibility of using a wave optical operator but does not provide a methodology. 
In addition to this work, a multi-beam optimization wave optical framework was recently introduced \cite{li2024multibeam}. The methodology and the optimization approach is similar to our work, but it is restricted to phase optimization and does not reveal details about the implementation. Their simulated volume is discretized with $N_\text{pixels}=256$  and $N_\text{angles}=64$. Hence, the simulated volume is around 10 times smaller and the total amount of optimized pixel values is 40 times smaller than in this work. Moreover, larger volumes and more beams is not possible in their implementation because of memory issues.

Here we introduce a scalar wave optical model which includes wave optical effects to TVAM that bears many similarities with diffraction tomography \cite{Wolf_1969, müller2016theory} which introduced the wave nature
of light to CT problems at micrometer resolution. Instead of tracing straight rays through the volume, we propagate the 2D projections step-wise through the volume with an angular spectrum method of plane waves (AS). This step is repeated for each angular projection pattern, and the intensities are then summed incoherently. 
This is similar to multi-slice diffraction tomography \cite{Ma_Xiao_Pan_2017} except that we do not take into account the variation of the refractive index during printing. 
Also we do not consider material properties such as the chemical diffusion of inhibitors; which other works have modeled computationally \cite{weisgraber2023virtualVAM}.
Based on efficient algorithms with custom gradient rules written in the programming language Julia\cite{bezanson2017julia}, we are able to simulate volumes as large as $550 \times 550 \times 550$ voxels with $600$ projection angles on a single NVIDIA A100 GPU. 
The comparison between wave optical and ray optical methods reveals that at \SI{20}{\micro\meter} the wave optical nature impacts the printing results noticeably.

\section{Theory}
\subsection{Angular spectrum of plane waves method}
    Free-space propagation allows for propagating vectorial components of the electrical field $\psi$  in a uniform medium with refractive index $n$ by solving the homogeneous Helmholtz equation \cite{Heintzmann_Loetgering_Wechsler_2023}
    \begin{equation}
        \nabla^2\psi + k^2 \psi = 0,
    \end{equation}
    where $k=\frac{2\pi}{\lambda_0 n}$ is the wave number. $\lambda_0$ is the wavelength in vacuum.
    If the field is known at a 2D plane $(x_0, y, z)$, the AS \cite{Goodman2017} allows to compute the field at a further location in space by
    \begin{equation}
        \psi(x_0 + x,y, z) = \mathcal{F}^{-1}[H_\text{AS}(x) \cdot \mathcal F[\psi(x_0,y,z)]] = \mathcal A[\psi(x_0,y,z)]
    \end{equation}
    where the kernel in Fourier space is given by
    \begin{equation}
        H_\text{AS}(x) = \exp\left(i 2\pi x\sqrt{\frac{1}{\lambda^2} - f_z^2 - f_y^2} \right)
    \end{equation}
    and the frequency coordinates are $(f_z, f_y) = \left(\frac{k_z}{2\pi},\frac{k_y}{2\pi} \right)$.
    Absorption is expressed via the absorption coefficient $\mu$ by the Beer Lambert's law at each slice of $\sqrt{\exp(-\mu \cdot (x - x_0))}$ . The $\sqrt{}$ is required if the attenuation is applied to the electrical field.
    The TVAM spatial light modulator projects a different pattern $\psi_\varphi(x_0, y, z)$ from each angle $\varphi$ into the volume
    in which the photoinitiator absorbs the light power. 
    $z$ is considered to be rotation axis of the vial. $x_0$ is the central slice of vial. So each pattern is imaged into the center of the vial.
    The defocus position $x$ is in the range $[-L/2, L/2]$ where $L$ is the side length of the volume.
    The full wave optical analog to the inverse Radon transform is given by:
    \begin{equation}
        \mathcal{P}_W\left[\psi_\varphi(x_0, y, z)\right](x',y',z) = \sum_{\varphi} \left|\mathcal R_{\varphi} \left[\mathcal{F}^{-1}[H_\text{AS}(x) \cdot \mathcal F[\psi_\varphi(x_0, y, z)]]\right](x',y',z)\right|^2
        \label{eq:main}
    \end{equation}
    where $\mathcal{R}_\varphi$ rotates the resulting 3D field in the $x,y$ coordinates into the correct direction from the illumination angle $\varphi$.
    Since the patterns are projected one after the other, they add incoherently. Note that in the wave optical optimization here, 
    we optimize amplitude patterns whereas ray tracing schemes operate with intensity patterns.

\subsection{Pattern optimization}
    Given the wave optical forward model, we can search for a set of patterns which minimizes a certain loss function for optimal prints.
    The loss function we use is inspired by \cite{Rackson_Champley_Toombs_Fong_Bansal_Taylor_Shusteff_McLeod_2021}.
    Additionally, we introduce an extra term to avoid overexposure of certain voxels.
    Mathematically, we seek a optimal set of patterns to minimize
        \begin{equation}
            \mathcal{L} = \underbrace{\sum_{v \,\in\,\text{object}} |\text{ReLu}(T_U - I_v)|^K}_\text{force object polymerization} + \underbrace{\sum_{v\,\notin\,\text{object}} |\text{ReLu}(I_v - T_L) |^K}_{\text{keep empty space unpolymerized}} + \underbrace{\sum_{v \,\in\,\text{object}} |\text{ReLu}(I_v - 1)|^K}_{\text{avoid overpolymerization}}
            \label{eq:loss}
        \end{equation}
  where $K$ can be 1 or 2 for $L_1$ or $L_2$ norm for example. $I_v$ is the normalized received intensity at a voxel $v$ after propagation of the patterns.
  The first term ensures that object voxel receive more intensity than $T_U$ to polymerize. The second term ensures that void voxels stay below a threshold of $T_L$ to not polymerize. The last term introduces a penalty if some object voxels receive more intensity than $1$ which avoids overpolymerization. Typical values are $(T_L, T_U) = (0.7, 0.8)$ and $K=2$.

\subsection{Implementation Details}
Pseudocode for the full wave optical optimization algorithm is sketched in Algorithm \autoref{alg}.
The basis is a gradient descent optimization scheme. Starting with a set of 2D amplitude patterns for each angle $\varphi$, the
patterns are propagated into space. After propagation of all angular patterns, the loss function given by \autoref{eq:loss} is calculated.
Then, the gradient of $\mathcal{L}$ with respect to the amplitude patterns $A$ is calculated.
With the gradient, we can perform one optimization step.
This optimization is restricted to amplitude optimization currently. Optimizing phase and amplitude at the same time is left for future work.
\begin{algorithm}
\caption{Gradient descent based wave optical optimization algorithm.}\label{alg}
\begin{algorithmic}[1]
\Require object, $M$, $T_L$, $T_U$, $\lambda$, $L$, angles
\State $A \gets$ array of size $(N,N,N_\text{angles})$ filled with zeros (initial guess) 
\State $i \gets 1$
\While{$i \leq M$}
    \State $A \gets \textrm{ReLu}(A)$ (map to non-negative) 
    \State $\psi_\varphi\gets$ $A \cdot \exp(0 \cdot i)$
    \State $I_\text{printed} \gets \mathcal{P}_W\left[\psi_\varphi(x_0, y, z)\right](x,y,z)$
    \State $\mathcal L \gets$ calculate loss function (see \autoref{eq:loss})
    \State $\nabla_A \mathcal L$ $\gets$ calculate gradient of $\mathcal L$ with respect to $A$
    \State $A \gets$ one optimizer step with gradient $\nabla_A \mathcal L$
    \State $i \gets i+1$
\EndWhile
\State $A \gets \textrm{ReLu}(A)$ (map to non-negative)\\
\Return $\psi_\varphi$
\end{algorithmic}
\end{algorithm}

\autoref{eq:main} is the key equation for the wave optical forward model. For the AS method, in fact we use the bandlimited AS \cite{Matsushima_Shimobaba_2009} implementation to prevent any wrap-around artifacts. 
The computational complexity to propagate a single 2D field $\psi_\varphi$ of size $(N \times N)$ into the 3D volume ($N$ different $x$ values) is given by $\mathcal{O}(N^3 \cdot \log N)$. This is because we propagate the field (size $N\times N$) $N$ steps into the volume and each of the FFTs evaluates at $\mathcal{O}(N^2 \cdot \log N)$.

\begin{algorithm}
\caption{Implementation of the wave optical forward model $\mathcal{P}_W$.}\label{alg2}
\begin{algorithmic}[1]
\Require patterns $\psi_\varphi$, $\lambda$,  $L$, $\text{angles}$, $N$
\State $I\gets$ array of size $(N,N,N)$ filled with zeros 
\State $i \gets 1$
\While{$i \leq \text{length(angles)}$}
    \State $\varphi \gets \text{angles}[i]$
    \State $\psi_\varphi(x,y,z)\gets$ $\mathcal A [\psi_\varphi(x_0, y,z)]$ (plane to volume)
    \State $I_\varphi(x,y,z)\gets |\psi_\varphi(x,y,z)|^2$
    \State $I_\varphi^R\gets \mathcal{R}_\varphi I_\varphi$
    \State $I \gets I + I_\varphi^R$
    \State $i \gets i+1$
\EndWhile\\
\Return $I$
\end{algorithmic}
\end{algorithm}

The amount of angles $N_\text{angles}$ covering $2\pi$ should be chosen approximately as $N_\text{angles} \approx N \cdot \pi$ (to avoid angular subsampling) which results in a total complexity of $\mathcal{P}_W$ equal to $\mathcal{O}(N^4 \cdot \log N)$.
This is promising because the Radon based optimization method has a computational complexity of $\mathcal{O}(N^4)$. 
It is obvious that a Radon based implementation will be faster in practice since the AS method in case of $\lambda \rightarrow 0$ evaluates to an identity matrix and is computationally trivial. 
Apart from $\log N$, the computational complexity is the same between wave optical and ray optical forward model. 

In Algorithm \autoref{alg2} we show the pseudocode for the full wave optical propagator. The most time consuming operation is line 5. First, a 2D FFT of the field is calculated and then in Fourier space the 3D kernel is multiplied and finally an inverse batched 2D FFT is calculated.
Each resulting volume is rotated with the rotation operator $\mathcal{R}_\varphi$ and added to the total intensity in the volume. 

Algorithm \autoref{alg} requires to adjoint the gradient of $\mathcal{P}_W$ with respect to the amplitude patterns $A$.
However, most reverse mode automatic differentiation (AD) engines such as the common ones in Julia (Zygote.jl \cite{Zygote.jl-2018}) or PyTorch \cite{pytorch} fail to calculate the gradient of the loss function in memory efficiently. 
In line 6 of Algorithm \autoref{alg2}, the pullback (backpropagator for the reverse mode AD) of $|\psi_\varphi(x,y,z)|^2$ is $2 \cdot \psi_\varphi(x,y,z) \cdot \Delta I_\varphi^{R}$.
In consequence, the AD needs not only to store the sum of all angular backprojections (memory consumption $\mathcal{O}(N^3)$) but also 
the result of each individual backprojection. 
This would require a memory of $\mathcal{O}(N^4)$. 
This memory scaling is beyond the memory limit of modern GPUs. For example, in our naïve PyTorch implementation volumes larger than $N\gtrsim100$ were not possible.
Instead of keeping all $\psi_\varphi(x,y,z)$ in memory, we  recalculate those fields in the gradient pass. This method is called memory checkpointing.
For the pseudocode of the whole adjoint of $\mathcal{P}_W$ see Algorithm \autoref{alg3}.

\begin{algorithm}
\caption{Adjoint of the wave optical forward model $\mathcal{P}_W$.}\label{alg3}
\begin{algorithmic}[1]
\Require $\Delta I$, $L$, $\lambda$, $\text{angles}$, $N$
\State $i \gets 1$
\While{$i \leq \text{length(angles)}$}
    \State $\varphi \gets \text{angles}[i]$
    \State $\Delta I_\varphi^{R}\gets \mathcal{R}_{\varphi}^* \Delta I$ (adjoint of rotation)
    \State $\Delta \psi_\varphi(x,y,z) \gets 2 \cdot \psi_\varphi(x,y,z) \cdot \Delta I_\varphi^{R}$ (recalculating $\psi_\varphi$ instead of storing it in memory)
    \State $\Delta \psi_\varphi(x_0,y,z)\gets \sum_{x} \mathcal{A}^*[\Delta \psi_\varphi(x, y,z)]$ (adjoint of $\mathcal{A}$) 
    \State $i \gets i+1$
\EndWhile\\
\Return $\Delta \psi(x_0, y,z)$ (gradients of each of the patterns for each angle $\varphi$)
\end{algorithmic}
\end{algorithm}

All algorithms were implemented in Julia and made use of \texttt{ChainRules.jl}\footnote{\url{https://github.com/JuliaDiff/ChainRules.jl}} to register a custom adjoint for the critical $\mathcal{P}_W$ operation.
Further, all routines feature CUDA acceleration in Julia \cite{Besard_Foket_De} and the bilinear differentiable image rotation routine is implemented in \texttt{KernelAbstractions.jl} \cite{churavy2021juliagpu} which results in fast CUDA executable code too.
As optimizer, we used the L-BFGS algorithm \cite{Liu_Nocedal_1989} implemented in \texttt{Optim.jl} \cite{Mogensen_Riseth_2018}.
L-BFGS is a quasi-Newton method which builds up an estimate of the inverse Hessian which helps minimizing the objective in less steps.
We encourage to use Fabio Crameris color-blind friendly color maps\cite{crameri_2023_8409685}.

\section{Simulation Results}
In this section we want to highlight three different simulations. Each of them was computed at $\lambda = \SI{405}{\nano\meter} / n$ where $n=1.5$ represents the refractive index of the resin. The discretization is $N=550$ with $N_\text{angles}=600$.
The absorption was set to 0, which implies that a set of angles in the range $[0, \pi)$ is sufficient.
To compare the wave optical optimization with the classical ray tracing approach, we optimize total volumes of sizes $L_1=\SI{100}{\micro\meter}$, 
$L_2 = \SI{1}{\milli\meter}$ and $L_3 = \SI{10}{\milli\meter}$. Each optimization was performed with Julia 1.10 on a NVIDIA A100 with 80GB GPU RAM. 
The runtime was roughly 6 hours for the wave optical optimization. The ray tracing optimization takes 10min. The L-BFGS optimizer requires usually around 50 iterations to achieve a well separated histogram as in \autoref{fig:general}f).
This work sets a computationally efficient and differentiable basis upon the rigorous pattern optimization can be based on.
\begin{figure}[h]
    \centering
    \includegraphics[width = \textwidth]{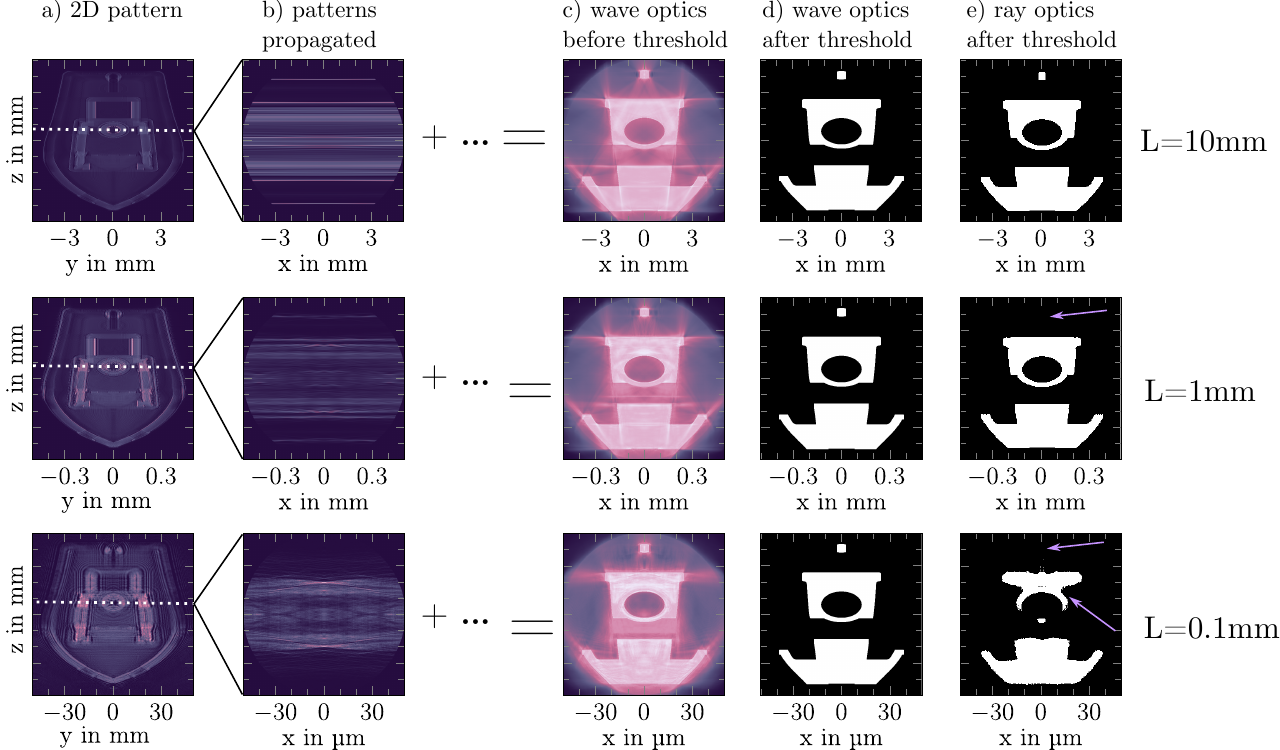}
    \caption{Overview of the simulation results for different printing sizes. a) is one 2D pattern. b) is the slice of this pattern propagated into the volume. c) indicates the incoherent sum of the patterns. d) after applying a relative energy threshold of 0.75. e) are the ray optical patterns propagated with the wave optical formalism and with the same threshold applied.}
    \label{fig:res}
\end{figure}

The results of the simulations are shown in \autoref{fig:res}. 
Each row represents a different resolution of the boat. The wave optical forward model results in the prints in column d). 
The Intersection over Union (IoU), a metric of fidelity between the target model and the resulting polymerized region, was always 1, so the print could be considered as perfect; in practice, absorbance from the photoinitiatior, chemical diffusion, and the time dependence of the refractive index upon polymerization would hinder resolution \cite{madrid2022controlling, Orth_Webber_Zhang_Sampson_de, rackson2022latent}. 
\begin{figure}[h]
    \centering
    \includegraphics[width = .5\textwidth]{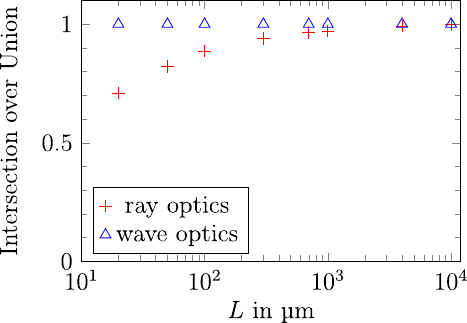}
    \caption{Intersection over Union for the different print results. The ray optical patterns are propagated with the wave optical model.}
    \label{fig:iou}
\end{figure}
In Figure 2e) we use the ray optical patterns as input for the wave optical forward model and apply the threshold of 0.75.
For $L_3=\SI{10}{\milli\meter}$ the wave optical and ray tracing optimization yield almost identical results. This is expected since
the wave optical nature is not dominating at those resolutions.
If we divide the physical dimension by 10, we see no substantial difference between the ray optical patterns and the wave optical patterns.
Only the small feature in the top (purple arrow) is not printed by the ray optical patterns.
However, at $L_1=\SI{100}{\micro\meter}$ the ray optical patterns start to misprint the object significantly. The wave optical patterns result in a perfect print.
In \autoref{fig:iou} we plot the general trend of the IoU for different resolution scales of the boat. The wave optical simulation was carried out for each datapoint. Especially for $L<\SI{200}{\micro\meter}$ the IoU drops significantly.
Quantifying the resolution limit of ray optical TVAM is not straightforward since it depends on the printed sample. 
But from this target we infer that below \SI{20}{\micro\meter} wave optical effects impact the print.

\section{Discussion}
This presented scalar wave optical modeling shows that in principle a amplitude modulated TVAM is able to print micrometer features inside a single photon absorptive resin.
Phase optimization only offers higher light efficiency but does not increase the resolution.
Our simulations also reveal that the recently printed \SI{20}{\micro\meter} features \cite{Toombs_Luitz_Cook_Jenne_Li_Rapp_Kotz-Helmer_Taylor_2022} are most likely limited by the ray optical assumptions underlying the existing optimization schemes. 
Of course, the next step is to experimentally demonstrate that wave optical patterns indeed allow for micrometer features.
This poses several challenges. At the moment, any refraction of the glass vial (or an index matching bath) is ignored. 
It remains unclear, how severe the modulation of the glass vial on the electrical field is. 
Describing strong refraction of glass interfaces with a wave optical model at millimeter scale is a computationally very demanding problem.
Further, the rotating vial in TVAM has to be digitally or mechanically stabilized to account for any motion blur.
Also, as recently demonstrated \cite{Orth_Webber_Zhang_Sampson_de}, chemical diffusion can have a major impact on resolution too.
Most likely, micrometer TVAM requires to take all those effects into account for reproducible printing.

\section{Backmatter}

\begin{backmatter}
\bmsection{Funding}
This project has received funding from the Swiss National Science Foundation under project number 196971 - “Light based Volumetric printing in scattering resins.”

\bmsection{Acknowledgments}
We would like to thank Simon Moser for discussion about the differentiable image rotation routine.
Also Baptiste Nicolet and Wenzel Jacob gave valuable input to the implementation of the optimization.
We appreciate the help of the Julia community regarding the implementation.

\bmsection{Disclosures}
Christophe Moser is a shareholder of Readily3D SA. All the other co-authors declare no conflict of interest.

\bmsection{Data Availability}
We strongly believe in sharing source code for physical simulations to verify correctness.
Hence, all simulation code is available over the \texttt{SwissVAMyKnife.jl} package on GitHub\footnote{\url{www.github.com/EPFL-LAPD/SwissVAMyKnife.jl}}.

\end{backmatter}

\bibliography{references.bib}

\end{document}